\documentclass[conference,letter,10pt,final]{IEEEtran}
\usepackage[utf8x]{inputenc}
\usepackage[T1]{fontenc}
\usepackage{graphicx}
\usepackage{longtable}
\usepackage{wrapfig}
\usepackage{rotating}
\usepackage[normalem]{ulem}
\usepackage{amsmath}
\usepackage{amssymb}
\usepackage{capt-of}
\usepackage{hyperref}
\usepackage[T1]{fontenc}
\usepackage{booktabs}
\date{}
\title{Optimization of a Radiofrequency Ablation FEM Application Using Parallel Sparse Solvers}
\begin{document}

\def\QRM{~QRMumps~}
\def\MAGMA{~MAGMA~}
\def\CUSOLVER{~cuSOLVER~}
\def\CUSOLVER_FMAD{~cuSOLVER_fmad_OFF~}
\def\MAGMA_LOW{~MAGMA_low_tol~}

\makeatletter
\let\orgtitle\@title
\makeatother

\title{\orgtitle}

\author{
\IEEEauthorblockN{
   Marcelo Cogo Miletto\IEEEauthorrefmark{1},
   Claudio Schepke\IEEEauthorrefmark{2},
   Lucas Mello Schnorr\IEEEauthorrefmark{1}}

\IEEEauthorblockA{\IEEEauthorrefmark{1}
Institute of Informatics, Federal University of Rio Grande do Sul, Porto Alegre, Brazil}

\IEEEauthorblockA{\IEEEauthorrefmark{2}
UNIPAMPA, Alegrete, Brazil}
}

\maketitle

\begin{abstract}
Finite element method applications are a common approach to simulate a
handful of phenomena but can take a lot of computing power, causing
elevated waiting time to produce precise results. The radiofrequency
ablation finite element method is an application to simulate the
medical procedure of radiofrequency ablation, a minimally invasive
liver cancer treatment. The application runs sequentially and can take
up to 20 hours of execution to generate 15 minutes of simulation
results. Most of this time arises from the need to solve a sparse
system of linear equations. In this work, we accelerate this
application by using three sparse solvers packages (MAGMA cuSOLVER,
and QRMumps), including direct and iterative methods over different
multicore and GPU architectures. We conducted a numerical result
analysis to access the solution quality provided by the distinct
solvers and their configurations, proposing the use of the peak
signal-to-noise ratio metric. We were able to reduce the application
execution time up to 40 times compared to the original sequential
version while keeping a similar numerical quality for the results.
\end{abstract}

\section{Introduction}
\label{sec:org6be9f38}

Scientific computing is a powerful domain to impel progress in many
fields of research. However, it is one of the most computing power
demanding activities, requiring a significant amount of computing time
to produce precise results in a feasible time. Many classical ideas,
such as the Finite Element Method (FEM), gained popularity through the
computational power support provided by modern hardware, enabling
scientists to deal with huge problems. FEM is nowadays widely used in
the industry and academy to simulate numerous phenomena. It is
described as a unified approach to various problems, and universally
adaptable to discrete systems \cite{zienkiewicz2005finite}.   

Such a standard computational method to tackle a handful of
problems is also widely studied. There are common steps in a FEM
application that are particularly interesting to look at when we are
concerned about the application performance. First, the global matrix
assembly step groups all element equations to represent the whole
problem. Then, obtaining the solution to this system of linear
equations represents the two most time-consuming steps. Furthermore,
the global matrix generally is very sparse since FEM is used to
approximating partial differential equations (PDEs), and PDEs are a
common source of sparse matrices \cite{saad2003iterative}. Therefore,
such applications cannot use basic factorization algorithms since the
fill-in overhead would be catastrophic for performance. Hence, such
problems demand the use of efficient sparse matrix solvers.  

The performance of sparse solvers, as well as for dense solvers, is
obtained through parallelizing the solution process. Numerous
libraries use different methods and approaches, exploring parallelism
in several computational resources using different programming
paradigms. Libraries such \texttt{MAGMA} \cite{tomov2010magma} and \texttt{CUSOLVER}
\cite{nvidiacusolver}, focus on providing a broad set of smaller
routines in a BLAS-like way, and complete solvers for both dense and
sparse matrices, containing direct and iterative solvers. On the other
hand, solvers like \texttt{QRMumps} \cite{agullo2013multifrontal}, and \texttt{PASTIX}
\cite{henon2002pastix}, focus on providing a unique, specialized sparse
direct solver for general sparse matrix and symmetric matrix.

The Radiofrequency Ablation FEM (RAFEM) \cite{jiang2010formulation}
simulates the Radiofrequency Ablation (RFA) procedure using the
FEM. Its original code runs sequentially, making the simulation
method very slow because of the computing-intensive steps involved
in FEM simulations. Furthermore, it uses the Frontal Method
\cite{irons1970frontal} for the solve step, which is a solution for
memory-efficiency but not for performance. Given these application
characteristics, the waiting time for simulating results can be
excessive, taking up to 20 hours to simulate 15 minutes of the RFA
procedure. Our work focus on the parallelization, numerical analysis,
and performance evaluation of the RAFEM application. We explore the
capacity of GPGPUs and multicore systems to parallelize and accelerate
the application, focusing on the solve steps, using multiple modern
libraries to obtain the equation system solution. The contributions
are as follows:  

\begin{itemize}
\item We implemented a parallel version of the RAFEM application that
works with multiple sparse solvers.

\item Tested and adapted the Fortran C interface of the \texttt{QRMumps}
application to a real application scenario.

\item We describe the parallelization methodology of the application using
GPU, and the role of sparse solvers, which can be valuable for other
FEM applications \cite{xu2016simulation,haemmerich2005automatic},
including other RFA  simulation software.

\item Conducted experiments in different machines with multiple solvers to
analyze the performance considering two workloads in the RAFEM
application context. We evaluate numerical results for each solver
used, comparing it to the original sequential version results.
\end{itemize}

The paper is organized as follows: Section \ref{sec:org1d978db} describes
the background concepts for this work, Section \ref{sec:orgd72cd31} describe
the methodological aspects and materials. The Section
\ref{sec:orgc8cbffb} present the analysis of the numerical results,
followed by Section \ref{sec:org01eb23b} that presents the
application performance analysis. Lastly, Section \ref{sec:orgbadce78}
presents the conclusion, recapping the interesting points, and
discussing future work. 

\section{Background}
\label{sec:org1d978db}
We describe the RAFEM application and its main loop in
\autoref{fig:execution_flow}, pinpointing the application execution
flow, its two most costly parts and the role of parallel sparse
solvers.

\subsection{The RAFEM application}
\label{sec:org2ede874}
RFA is a minimally invasive medical procedure. The RAFEM application
simulates the RFA for treating hepatic cancer. The procedure consists
of using the Joule heating energy from a special electrode connected
to an alternated current generator to eliminate cancer cells by
heating the desired liver region. After the positioning of the
electrodes inside the patient's liver, when the RFA starts, techniques
like computed tomography or magnetic resonance imaging are unable to
provide precise observations to inspect the temperature distribution
area during the actual procedure \cite{jiang2010formulation}. 

To the procedure be successful, the heated area in the liver must
eliminate all cancer cells. Otherwise, tumor recidivism can occur. To
understand and adjust the procedure parameters for each treatment
case, specialists must know, beforehand, the total heated area in the
liver. Thus, simulation can increase the chances of success of the
procedure in avoiding resubmitting the patient to a new RFA
procedure. The RAFEM application development was to help clinician
specialists find the best parameters for each treatment case through
computer simulation. The application uses the FEM to simulate the heat
and voltage distribution along time in a mesh of tetrahedral
elements. 

\begin{figure}[!htbp]
  \centering
  \includegraphics[width=.48\textwidth]{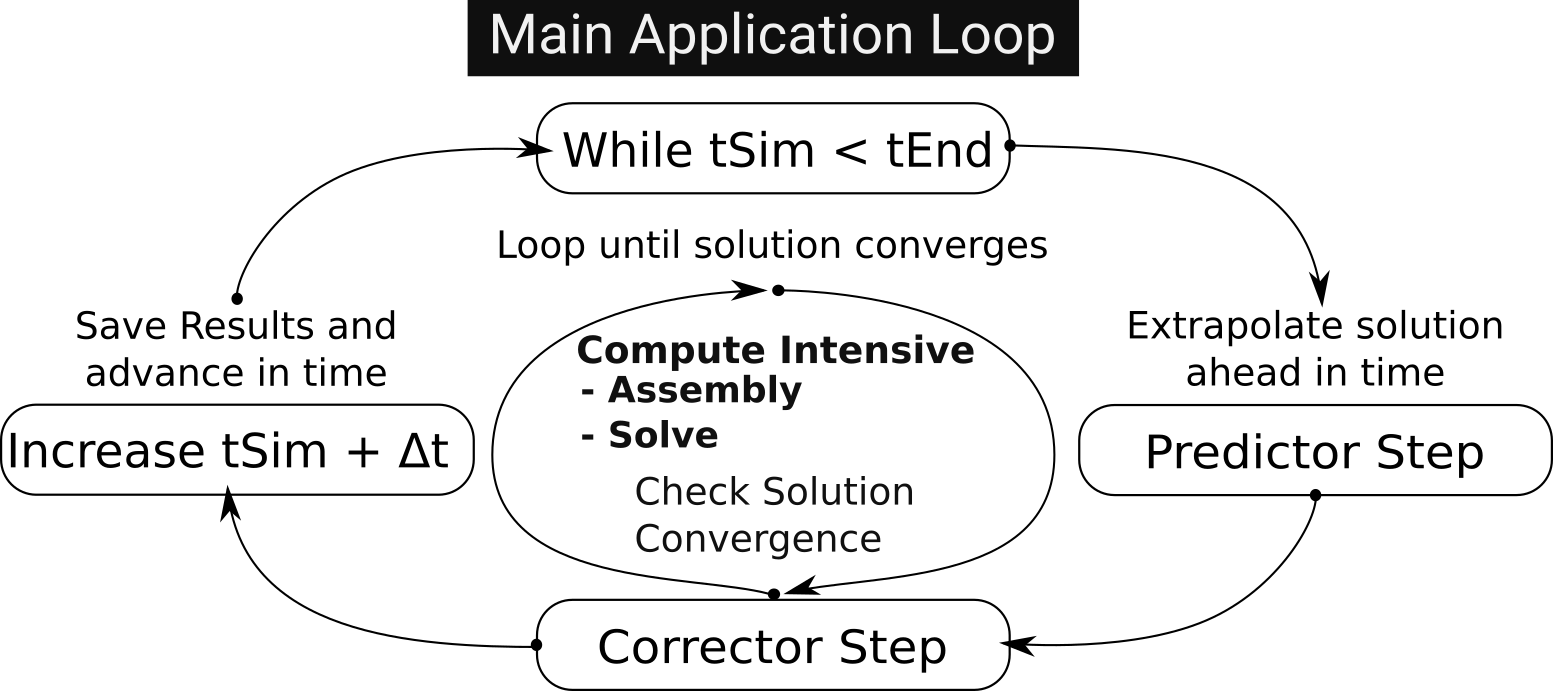}
  \caption{Main loop of the RAFEM application.} 
  \label{fig:execution_flow} 
\end{figure}

\textbf{Main application loop:} The application uses the predictor-corrector
method to extrapolate the solution further in time, then correct it
until reaching a convergence threshold value. The
\autoref{fig:execution_flow} depicts the principal application
execution flow. After the pre-processing part of reading data and
setting initial values for the problem, it enters a loop that controls
the total simulated time, advancing in dynamically sized time steps of
size \(\Delta t\) \cite{jiang2010formulation}. The loop performs \(n\) steps,
which depends on the total simulation time. For each one of these
steps, it executes the corrector step until the desired convergence
threshold is satisfied. For this corrector step, the application
computes all the element equations with updated values and assembly
them in the global matrix that is structurally symmetric and sparse,
which needs to be solved by a sparse matrix solver. This corrector
step is the most costly in the application. It runs many times and
contains the two most time-consuming phases of FEM applications: the
global matrix assembly and solve.

\textbf{Assembly step:} Many advances focus on these steps towards exploring
the use of GPUs to accelerate them \cite{georgescu2013gpu}. For the
assembly step, including numerical integration, a variety of works
study different approaches, focusing on accelerating this step using
GPUs. Approaches consider using better formats to represent and
assembly the global matrix besides the most common ones, like the
proposed adapted coordinate (COO)  and the ELLpack (ELL) format
\cite{amorim2020gpu,sanfui2017twokernel}. Furthermore, designing
specialized algorithms and data structures aligned to GPU programming
specific concepts like dynamic GPU memory allocation, thread-block
shared memory, and warps \cite{amorim2020gpu,kiran2018warp} seems to be
a common approach. The authors discuss the advantages and
disadvantages of using techniques like graph coloring, graph
partitioning, atomic operations, and precomputed structures to handle
race conditions. For which there is not a consensus on what is the
best strategy. The best approach depends on the problem structure and
mesh type, computational capability of the devices, and the FEM
parallelization approach. For example, despite those works that use
GPU to accelerate the global matrix assembly in a sparse matrix
format, some GPU FEM problems are modeled in the Element-by-Element
FEM (EbE-FEM) approach \cite{EbE2012CUDA,martinez2015efficient}, which
avoids explicitly assembling the global matrix to obtain its
solution. 

\textbf{Solve Step}: For the solve phase, the advances are essentially
expressed by the wide variety of parallel iterative and direct sparse
solvers that we have available today. The solvers use different
approaches, like in the GPU-specific libraries that have collections
of parallel solvers that fit in many cases
\cite{nvidiacusolver,tomov2010magma}. Besides these libraries and their
comprehensive set of routines, other approaches focus on developing
specialized tunned algorithms. These specialized algorithms are often
used for sparse direct solvers because they are more challenging in 
the aspects of parallelization, communication, and scalability. The
MUMPS \cite{amestoy2000mumps} solver is one of the earliest distributed
memory parallel sparse solvers, capable of distributing and balancing
the work among many cores. MUMPS achieved such capability through the
dynamic scheduling of computational tasks, manually implemented and
extracted from the algorithm's knowledge. Nowadays, many
high-performance direct sparse solvers follow a task-based approach,
looking to achieve performance scalability and ease in managing the
tasks. Solvers like QRMumps \cite{agullo2013multifrontal}, and
Pastix \cite{henon2002pastix} are examples of solvers that rely on a task-based
approach, allowing them to use powerful scheduling algorithms and the
possibility to explore heterogeneous computational resources. There
are also parallel GPU solvers designed especially for EbE-FEM
problems \cite{yan2017research}. 

\subsection{Parallel Sparse Solvers}
\label{sec:org7a77653}

Consider the problem of solving a linear system of equations in the
form of \(Ax = b\).  Essentially, there are two families of solvers
\cite{tu2018computational}: the direct solvers and the iterative
solvers. These two kinds of methods fit for solving both sparse and
dense systems of linear equations. The major difference between these
approaches is that direct methods use a finite number of elemental
operations to factorize the input matrix \(A\) into a form that
obtaining the solution becomes trivial. Iterative solvers obtain the
solution by performing a sequence of approximations of the solution
using matrix-vector multiplications. The cost of such methods can be
said to be \(IN^2\), where \(I\) is the number of necessary iterations to
obtain a solution that is sufficiently close to the real solution
given a tolerance value. While iterative methods naturally fit for
sparse problems because they do not change matrix structure, direct
solvers must use specialized algorithms and data structures to obtain
advantage from the matrix sparsity.

Some methods utilization depends on the particular matrix
properties. Thus, the type of solver and which algorithm to use is
dependent on the type of problem matrix. For symmetric
positive-definite matrices, for example, the direct method of Cholesky
factorization and the Conjugate Gradient iterative method are very
common choices. However, when the matrix properties do not hold, we
must use other methods like the Restarted Generalized Minimal Residual
Method GMRES(m) and the sparse QR factorization. The cost of direct
methods for sparse factorization depends much on the specific
algorithm implemented and the matrix sparsity pattern. These factors
have a significant influence on the total fill-in introduced during
factorization. For this, sparse factorization is commonly preceded by
an analysis step. This analysis step is responsible for applying
fill-reducing permutations in the original matrix such as the Reverse
Cuthill-McKee, Average Minimum Degree, and Nested Dissection
reorderings. The permutation is commonly followed by a symbolic
factorization step to compute the final matrix structure. Analogous to
this analysis step in direct sparse factorization. Iterative solvers
may suffer from a low convergence rate when we have big sparse
problems or ill-conditioned matrices. Preconditioners are applied to
alleviate this effect, accelerating the solver convergence rate.

\section{Methods and Materials}
\label{sec:orgd72cd31}
\begin{figure}[!htbp]
  \centering
  \includegraphics[width=.5\textwidth]{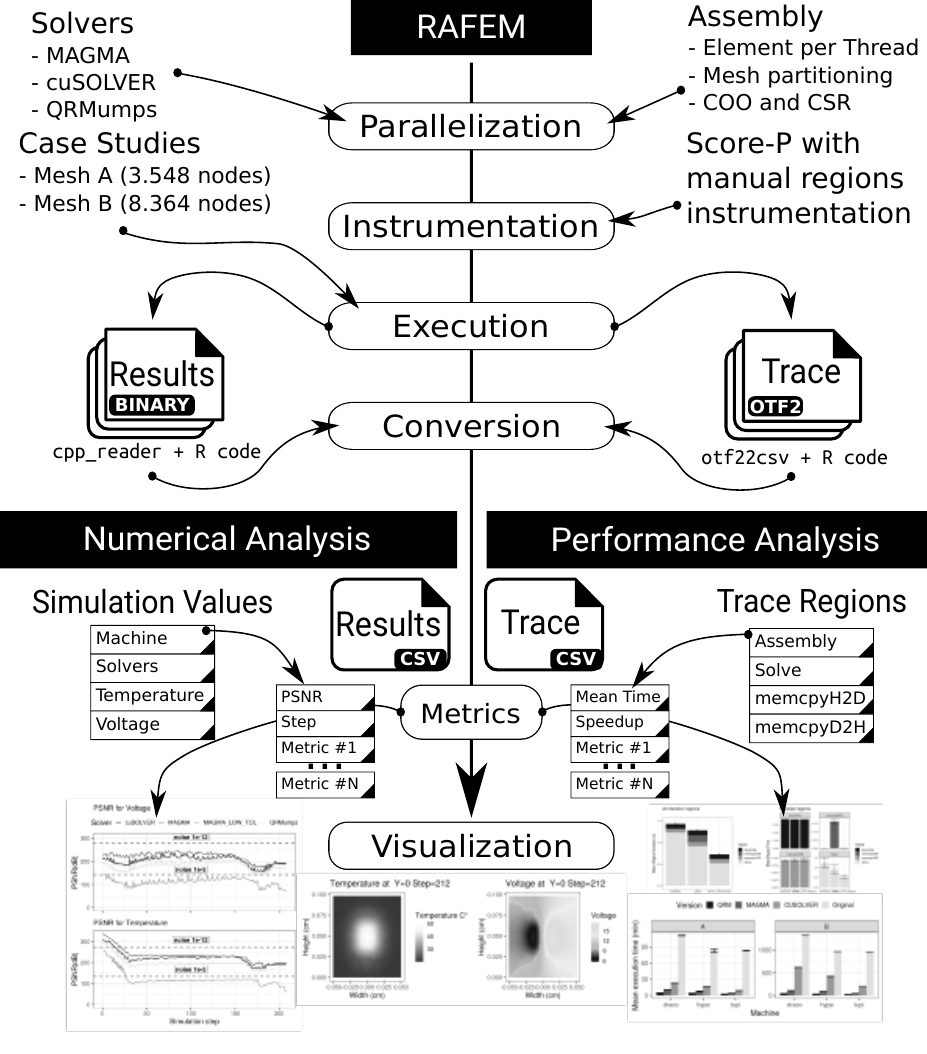}
  \caption{Methodological workflow overview from application parallelization 
  and instrumentation, to data collection and visual presentation.} 
  \label{fig:methodological_workflow}
\end{figure}

We present the methodological workflow of this paper, as depicted in
\autoref{fig:methodological_workflow}. The workflow starts with the
data collection, involving the parallelization and instrumentation,
and the execution of our experiments. Then, executions can generate
two types of results: the ScoreP OTF2 trace results and the simulated
results written to a binary file. These raw files need to be
post-processed in the conversion step, enabling us to work with the
data files to derive metrics and create visual representations for
numerical and performance analysis.

\subsection{Parallelizing Application}
\label{sec:orgca5b46e}

The assembly stage parallelization was done in a previous work
\cite{rafemPDP2020} by implementing CUDA kernels that were responsible
for assembling the individual elements equations and grouping them in
the Compressed Sparse Row (CSR) sparse matrix format to represent the
global matrix. Despite all the promising techniques for the assembly
phase presented in Section \ref{sec:org1d978db}, we kept this simpler
approach of assembling the element matrices individually (one thread
by element) and assigning them in the global matrix by mapping threads
to assembly specific matrix regions, similar to a graph partitioning
approach to avoid race conditions. Thus, there is a potential
improvement to explore in the assembly part. However, we focus on the
solve phase in the RAFEM application as the most significant
computational part arises from it.

The previous work already used \texttt{MAGMA} for the solution phase but did
not explore different solvers and computational environments. Besides
the \texttt{MAGMA} solution, we analyze two more solvers in this work: the
sparse QR factorization from \texttt{cuSOLVER}, done entirely in GPU, and the
task-based QR sparse factorization from \texttt{QRMumps}, using CPUs. The
previous code assembly the matrix in the CSR format, which fits for
\texttt{MAGMA} and cuSPARSE solver methods. However, \texttt{QRMumps} expects a matrix
in the COO format as input. Thus we extended the CUDA kernel that does
the matrix assembly in CSR format to create the COO rows
representation array for the \texttt{QRMumps} solver. We employ the \texttt{QRMumps}
C-Fortran interface to include the solver in the application. 

For the \texttt{QRMumps} solver, in every corrector step iteration (innermost
loop in \autoref{fig:execution_flow}), we make an extra memory copy of
the global matrix sparse representation from the GPU to the
CPU. Moreover, a copy of the solution needs to be in the GPU for the
next assembly step. These copies are unnecessary for the GPU solvers
since the solution is already in the GPU. However, we need to copy the
solution back to CPU every corrector iteration for the GPU solution
phase. This last copy is needed because the application performs the
predictor step, the convergence check of the corrector step, and the
time increment calculation sequentially as the original code
do. Nonetheless, the total execution time impact difference implied by
these extra copies is neglectable (below 0,15\%). Lastly, we
parallelize seven out of eleven vector reduction operations and
transformations that arise from the prediction, assembly, and solve
steps using the \texttt{thrust} library and its operations \texttt{transform} and
\texttt{transform\_reduce}.

\subsection{Solver Tuning}
\label{sec:orga1ec5cd}

Having the three solvers working, we started to study parameters that
affect their performance. The matrix reordering algorithm is crucial
for direct solvers to enhance parallelism and control the fill-in
effect. For \texttt{QRMumps}, we explored the Scotch
\cite{pellegrini1994static} and Metis \cite{karypis1998fast} reordering
algorithms. Metis provided better acceleration for the matrix
factorization, which was the best reordering for \texttt{cuSOLVER} QR as
well. Regarding the \texttt{QRMumps} application, as it performs a tiled
Householder QR factorization, we need to set the dimensions of the
blocks and internal blocks. We have chosen the values of 320 and 32,
respectively, which are typical values for this type of
configuration. Furthermore, We compiled \texttt{QRMumps} with the StarPU
1.3.3 \cite{augonnet2011starpu}, a task-based library that handles the
tasks creations and distribution among computational resources. We
configured the solver scheduler to use the local-work stealing (LWS)
strategy, which presents good performance for CPU-only environments
\cite{thibault2018runtime}.    

Studying the problem, we noticed that the matrix structure remains the
same, which allows us to reuse the column permutation done in a
previous step. By reusing the reordering, we save time in the analysis
phase that precedes the factorization. Unfortunately, the \texttt{cuSOLVER}
function call does not enable passing a user-defined reordering as
\texttt{QRMumps} do. To contribute to the flexibility of the solver, we made a
patch for \texttt{QRMumps} C-Fortran interface. This patch enables the user to
retrieve the array containing the column reordering made by the chosen
reordering algorithm. With access to this array, we can reuse it in
later iterations of the solve phase for the \texttt{QRMumps} application. 

\begin{figure}[!htbp]
  \centering
  \includegraphics[width=.48\textwidth]{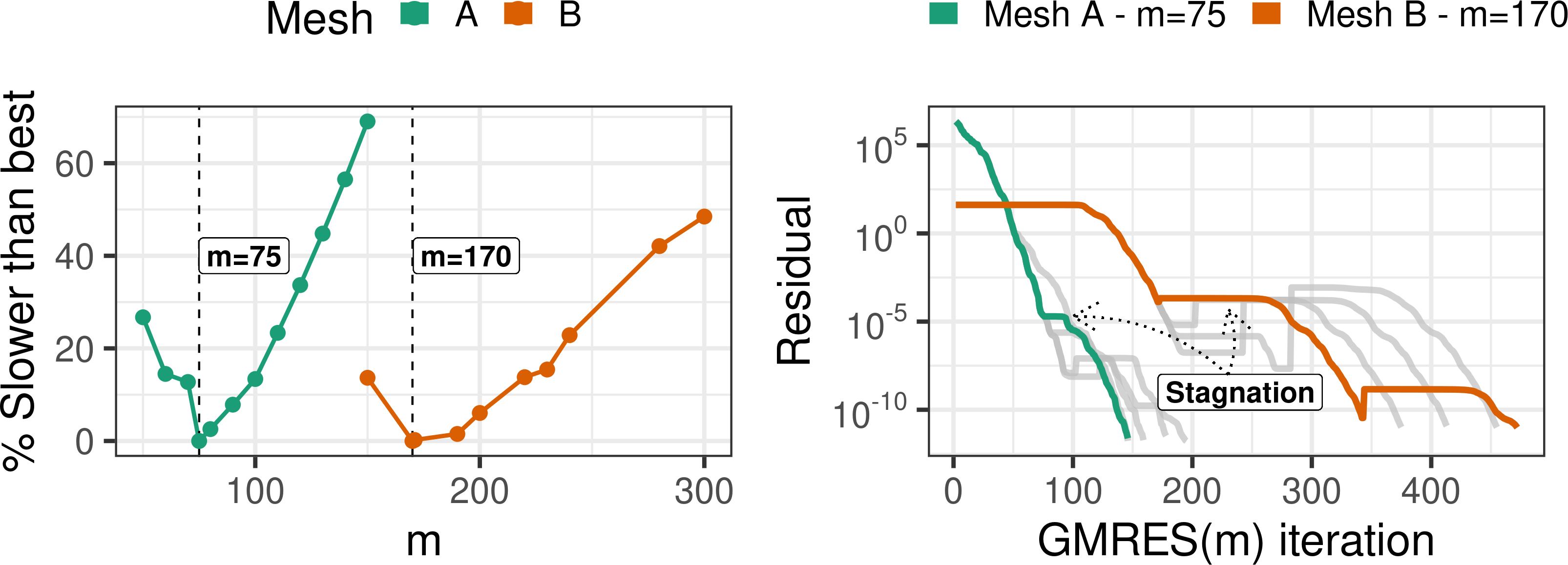}
  \caption{Best value for m in each mesh (left), and stagnation effect (right).} 
  \label{fig:find_gmres_m}
\end{figure}

Lastly, the \texttt{MAGMA} library method GMRES(\(m\)) has the parameter \(m\)
that impacts the solver convergence rate, which directly affects its
performance by increasing or decreasing the number of needed
iterations to reach the desired tolerance for the residual
value. Recognizing this effect, we did some previous experiments to
find a fair value of \(m\) for our tested workloads, as presented in
\autoref{fig:find_gmres_m}. In the left part of the figure, we have
the \(m\) values in the x-axis, and how many times the total application
execution time was slower than the smaller execution time obtained for
that value of \(m\). We could achieve further improvements in the
\texttt{MAGMA} GMRES(m) method efficiency if it incorporates some ideas, such
as using an adaptative \(m\) value during the simulation and for each
solution \cite{cabral2020improving}. Such improvements can reduce
stagnation effects observed in the right part of the
\autoref{fig:find_gmres_m}, which occurs even for the best-picked
values for some iterations of the solve phase. Moreover, we set the
convergence tolerance for the GMRES(\(m\)) method to \(10^{-10}\) in our
experiments.

\subsection{Tracing Code Regions}
\label{sec:orge8fa0b0}

A common approach to improving application performance is first
understanding the application behavior to identify possible
performance bottlenecks and improvement points. A popular way to do it
is by application tracing and profiling. The ScoreP \cite{an2011score}
library provides a simple user-defined tracing, allowing the user to
manually define tracing of code regions areas that the user judges
interesting. These regions tracing help collect simple measurements in
application-specific points without generating too much overhead in
application execution time caused by tracing intrusion
\cite{an2011score}. 

We focused on tracing the costly computing regions of the corrector
step, including the assembly and solve parts and all the data
movements between CPU and GPU. This tracing enables us to characterize
the computational cost of this part of the application in four
different common regions: Assembly, Solve, memcpyH2D (memory copies
from the host (CPU) to the device (GPU)), and memcpyD2H (from device
to host). Thus, we have a detailed view of the application
performance, which can help in the decision of optimizing one of these
parts and predicting how much we can reduce the total execution time
by its optimization. The ScoreP tool records application trace in the
OTF2 format. We used a conversion tool named \texttt{otf22csv}\footnote{\url{https://github.com/schnorr/otf2utils}} to
transform the OTF2 trace data into the CSV format, enabling the
analysis with modern data analysis tools like the R language.   

\subsection{Experimental Context and Workload Detail}
\label{sec:org75337d4}

We used machines with different hardware capabilities to evaluate the
performance of the studied solvers. The machines description is listed
in Table \ref{tab:orga6c9844}. All of them used CUDA version 10.2.89, \texttt{MAGMA}
version 2.5.2, and StarPU 1.3.3 and openblas 0.3.9 for \texttt{QRMumps}. The
major difference between them is the number of CPU cores and the GPU
specifications. Table \ref{tab:org4fd03f4} presents the two different
resolutions of finite element meshes that describe the liver and the
electrodes used as workloads. Both meshes A and B produce square and
very sparse matrices. To obtain metrics to calculate mean execution
time and speedup, we used a full factorial design considering the
three factors \texttt{machine(3)}, \texttt{workload(2)}, and \texttt{solver(4)}. Making 30
repetitions for the smaller workload A and 10 for the bigger one B. In
total, we performed 480 executions to obtain the performance metrics.   

\begin{table}[htbp]
\caption{\label{tab:orga6c9844}Hardware information of machines used in the experiments.}
\centering
\scriptsize
\begin{tabular}{llrr}
\toprule
\textbf{Machine} & \textbf{CPU} & \textbf{GPU} & \textbf{CUDA cores}\\[0pt]
\midrule
Draco & 2\texttimes{}8 E5-2640v2 - 2.0 GHz & Tesla K20 & 2.496 - 706 MHz\\[0pt]
Hype & 2\texttimes{}10 E5-2650v3 - 2.3 GHz & Tesla K80 & 4.992 - 824 MHz\\[0pt]
Tupi & 1\texttimes{}8 E5-2620v4 - 2.1 GHz & GTX 1080Ti & 3.584 - 1.582 MHz\\[0pt]
\bottomrule
\end{tabular}
\end{table}

\begin{table}[htbp]
\caption{\label{tab:org4fd03f4}Workload information used in the experiments.}
\centering
\scriptsize
\begin{tabular}{llllrr}
\toprule
\textbf{Mesh} & \textbf{\# Nodes} & \textbf{\# Elements} & \textbf{Matrix size} & \textbf{Nonzeroes} & \textbf{Sparsity}\\[0pt]
\midrule
A & 3.548 & 18.363 & 7.096\texttimes{}7.096 & 189.462 & 0.0037\%\\[0pt]
B & 8.364 & 44.811 & 16.728\texttimes{}16.728 & 450.451 & 0.0016\%\\[0pt]
\bottomrule
\end{tabular}
\end{table}

\section{Numerical Results Validation}
\label{sec:orgc8cbffb}
Most important than accelerating the application, is to check if the
parallel versions still producing valid numerical results. As there is
no such thing as infinite precision in floating-point numbers, we are
limited to the machine epsilon or machine precision. This precision
limit can cause roundoff errors that can be magnified or propagated
during the application execution. Another aspect that needs attention
when evaluating precision is that we are running operations in
parallel so that the rounding order may produce slightly different
results. Lastly, operations like the Fused Multiply-Add (FMA), present
in NVIDIA GPUs, perform operations of type \(X×Y+Z\) using a single
rounding step. Because of this, the computation of the results is more
accurate \cite{whitehead2011precision}. 

To evaluate and quantify the quality of our numerical solution
obtained by the multiple solvers used, we propose using the Peak
Signal-To-Noise Ratio (PSNR) \cite{Korhonen2012PSNR}, a common metric
used for quality estimation involving images. We compare the binary
written values for the temperature and voltage along simulation steps
produced by RAFEM. We used the original sequential version results as
the values that remain unaffected by noise. Then, we compare each step
solution using the following formula for calculating PSNR: 
\begin{equation}
PSNR_i = 20 \times log_{10} \times \frac{MAX(A)}{\sqrt{MSE_i}},
\end{equation}
where the Mean Squared Error (MSE) is
\begin{equation}
MSE_i = \frac{1}{n} \times \displaystyle\sum_{j=1}^{n}[A_{ij} - B_{ij}]^2
\end{equation}

\begin{figure}[!htbp]
  \centering
  \includegraphics[width=.47\textwidth]{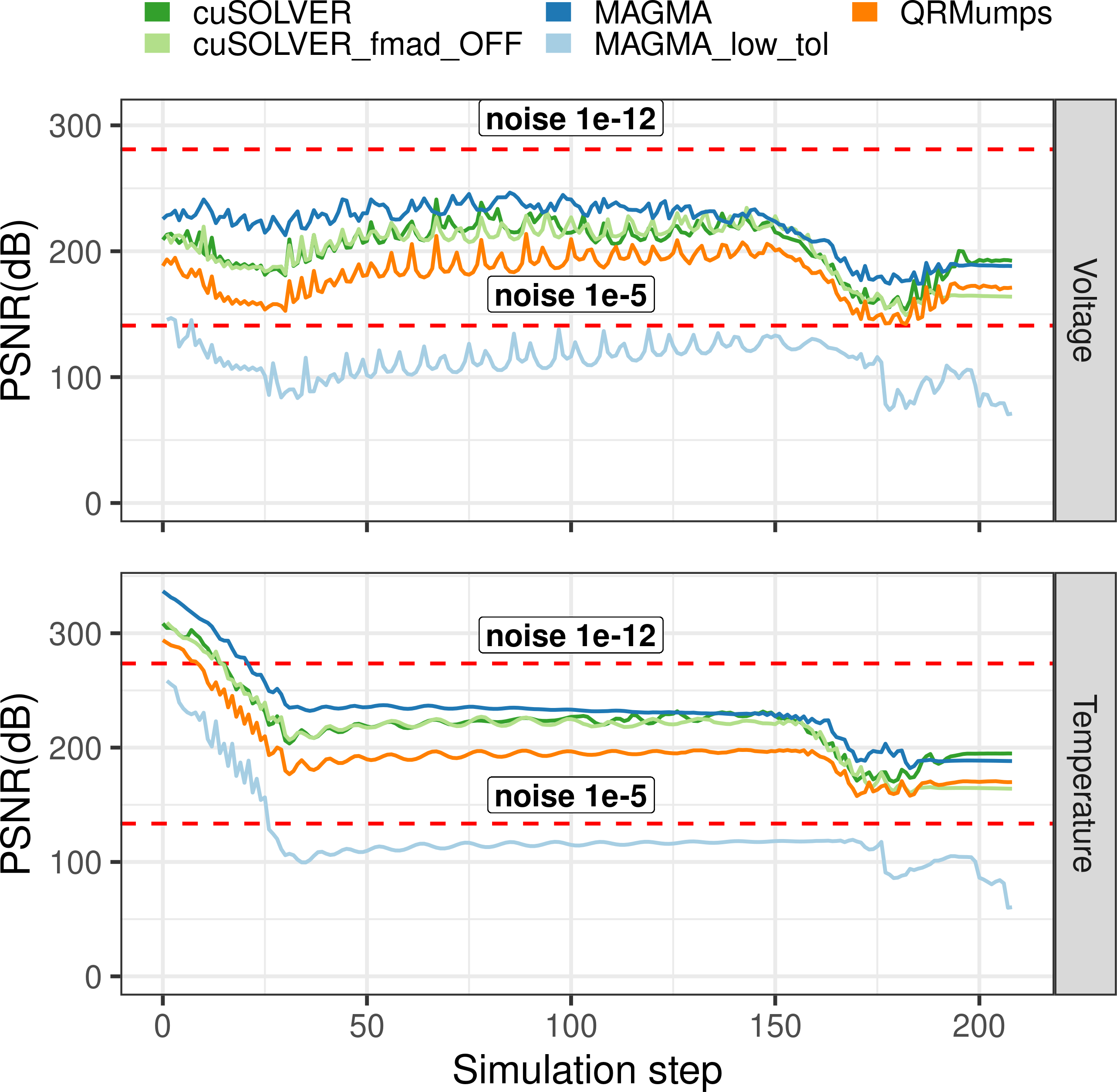}
  \caption{PSNR for voltage and temperature for mesh A.} 
  \label{fig:PSNR_meshA}
\end{figure}

\begin{figure}[!htbp]
  \centering
  \includegraphics[width=.47\textwidth]{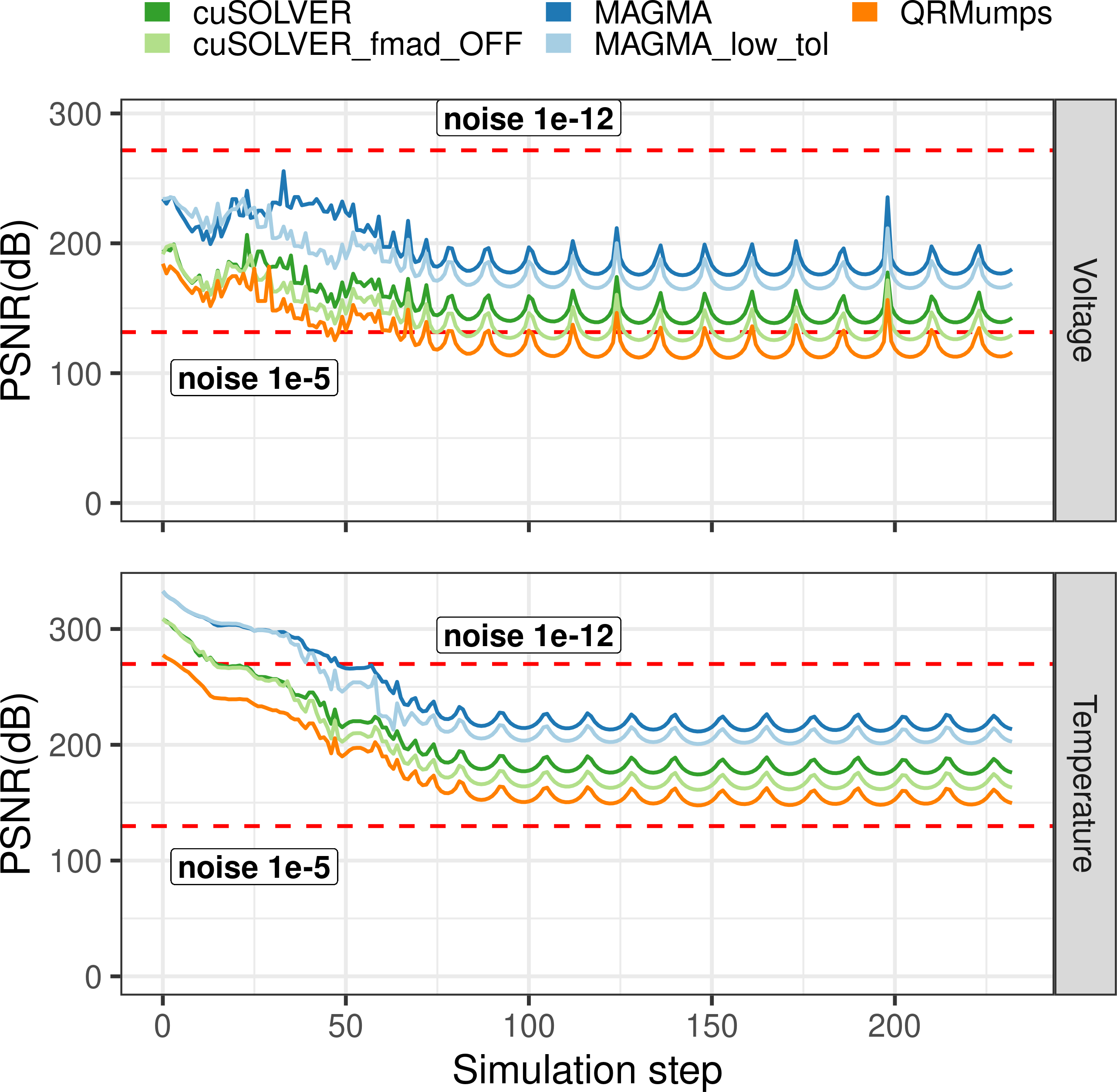}
  \caption{PSNR for voltage and temperature for mesh B.} 
  \label{fig:PSNR_meshB}
\end{figure}

The \(PSNR_{i}\) value represents the PSNR value for simulation step \(i\),
\(MAX(A)\) is the maximum observed value for temperature or voltage
among all steps, \(n\) is the number of nodes of the mesh, and \(A_{i}\) and
\(B_{i}\) are the original and some parallel version solution for all nodes
\(j\) in step \(i\). Accordingly, we defined a PSNR value for each of the
simulation steps to generate the Figures \ref{fig:PSNR_meshA} and
\ref{fig:PSNR_meshB}. In these figures, we included different
configurations for \texttt{MAGMA} and \texttt{cuSOLVER} to observe their impact on this
metric. The \texttt{MAGMA\_low\_tol} is the \texttt{MAGMA} GMRES(\(m\)) solver with a
tolerance of \(10^{-6}\) compared to the \texttt{MAGMA} that uses a tolerance of
\(10^{-10}\). The \texttt{cuSOLVER\_fmad\_OFF} is the \texttt{cuSOLVER} version compiled with
the flag \texttt{-{}-{}fmad=false}, which disables the GPU's FMA operation. Lastly,
the red dashed lines present control values to know what PSNR value we
have when calculating it using the original solution values summed
with a noise of \(10^{-5}\) and \(10^{-12}\) in each mesh node.

We observe that despite the \texttt{MAGMA\_low\_tol} PSNR value for mesh A
(\autoref{fig:PSNR_meshA}), the \texttt{QRMumps} solver presented a lower value
for PSNR, indicating higher differences in the numerical results. In
contrast, \texttt{cuSOLVER}, which also performs the QR sparse factorization,
presented a higher value for PSNR. Even when FMA disabled
(\texttt{cuSOLVER\_fmad\_OFF}), except in the last steps for mesh A
(\autoref{fig:PSNR_meshA}). The other interesting result that we
observe is how the FMA operation changes the PSNR values a bit for
\texttt{cuSOLVER}, mainly in the B mesh (\autoref{fig:PSNR_meshB}). 

For the \texttt{MAGMA\_low\_tol}, we can observe in \autoref{fig:PSNR_meshA} that
it presents a much lower PSNR value than the others. We also noticed
in this case that the corrector step did not converge in less than the
maximum allowed iterations (50) for step number 208. When this occurs,
the RAFEM application does not accept the solution for that step and
reduces the time step size in half. So there is a mismatch in time
after that, but the last step represents the same final time, which is
the total simulated time, set to 15 minutes for all
experiments. However, for the bigger case B in
\autoref{fig:PSNR_meshB}, the \texttt{MAGMA\_low\_tol} presented a higher PSNR
than other solvers, being just below the \texttt{MAGMA} solution with a
tolerance of \(10^{-10}\). In the overall view, this metric shows us that
different solvers and configurations provide slightly different
numerical results. However, their increase and decrease behavior for
the PSNR value is very similar, except for the case where the solution
correction at step 208 does not converged for mesh A.

\begin{figure}[!htbp]
  \centering
  \includegraphics[width=.49\textwidth]{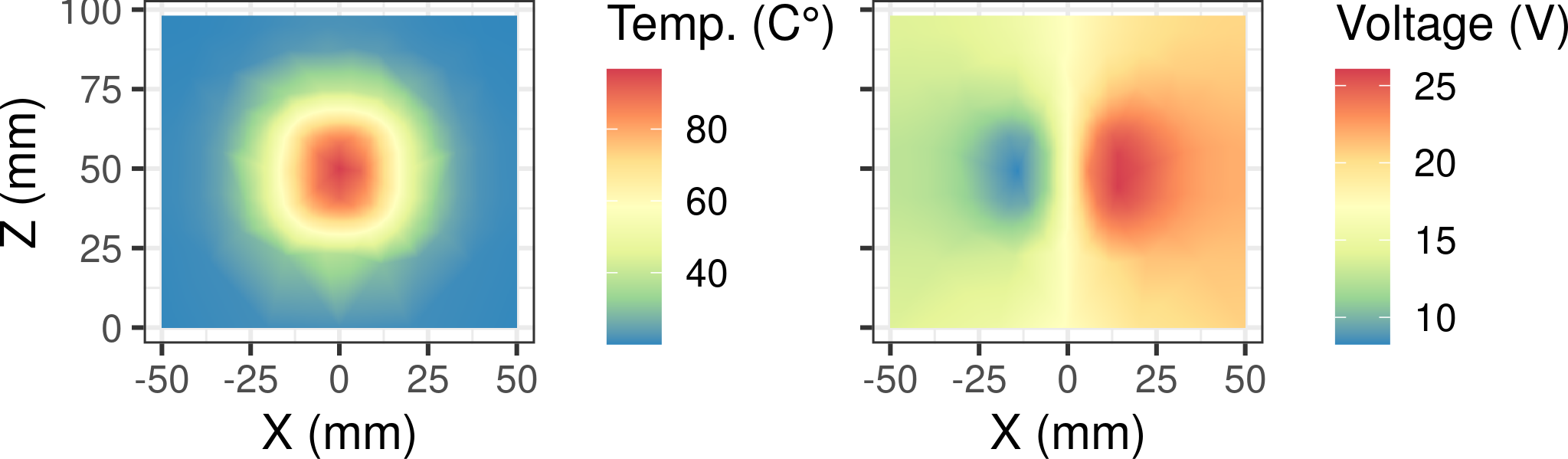}
  \caption{Original results for step 177 with the electrodes side-by-side.
  Distribution of temperature (left), and voltage (right).} 
  \label{fig:original_results}
\end{figure}

One drawback of using the PSNR metric is that it summarizes the
distribution of the numerical differences, hiding them in this
metric. For example, PSNR is incapable of telling us if there are many
small differences or fewer more significant differences for the
simulated values at a given step. Furthermore, we are also unable to
observe the spatial dispersion of these differences. Hence, we used
the binary file results to generate 2D surface plots of the step that
presented lower PSNR value (step 177 for mesh A), presenting the
difference of the values from the original solution in a spatial
way. We fixed the Y value as 0, which cuts the two electrodes of the
mesh in the middle, providing a view of them
side-by-side. \autoref{fig:original_results} presents the original
values for temperature and voltage at step 177 for the same case
execution of \autoref{fig:PSNR_meshA}. We observe that the temperature 
spread is symmetrical as well as the voltage, but with lower values in
one side to create the current flow between the electrodes. We used
the \texttt{akima} \cite{akima2016akima} \texttt{R}  package to create the 2D
interpolation of the values. 

\begin{figure}[!htbp]
  \centering
  \includegraphics[width=.49\textwidth]{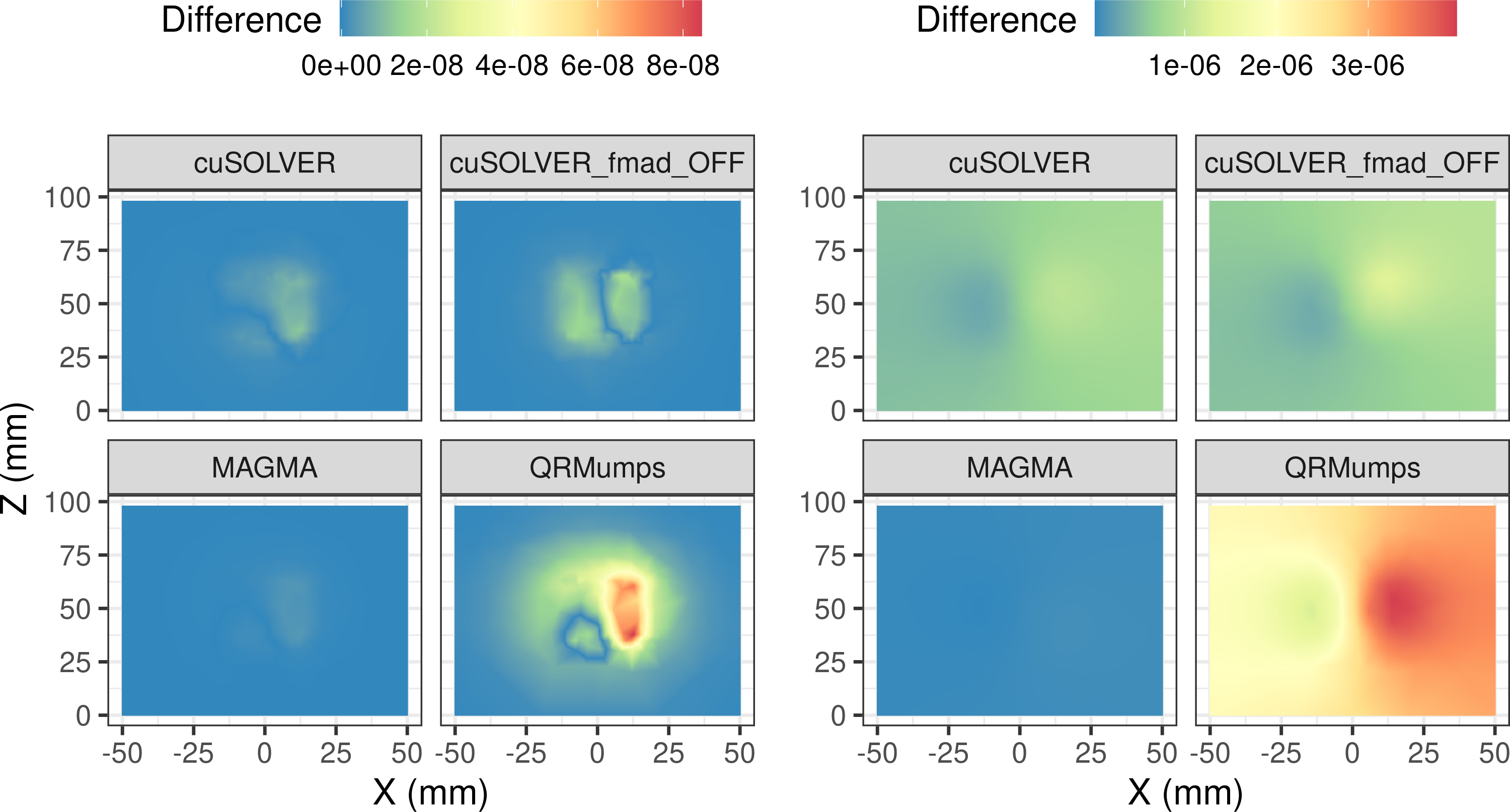}
  \caption{Numerical differences of step 177 for temperature (left) and for voltage (right), facing the electrodes side-by-side ($Y$ = 0).}  
  \label{fig:differences_results}
\end{figure}

The \autoref{fig:differences_results} presents the visualization of
the absolute differences, showing the temperature on the left side,
the voltage on the right, and facetting plots by the solvers. Note
that the difference in color scale and value magnitude is very
different between temperature and voltage. In general, we observe that
the area between X [\(-25\), \(25\)] and Z [\(25\), \(75\)] holds most of the
numerical differences, which is the area near the position of the
electrodes. These figures enable us to see that no significant
numerical differences have arisen in a spatial sense, spreading only
around that area and did not appear in the edges, for example. 

\begin{figure}[!htbp]
  \centering
  \includegraphics[width=.49\textwidth]{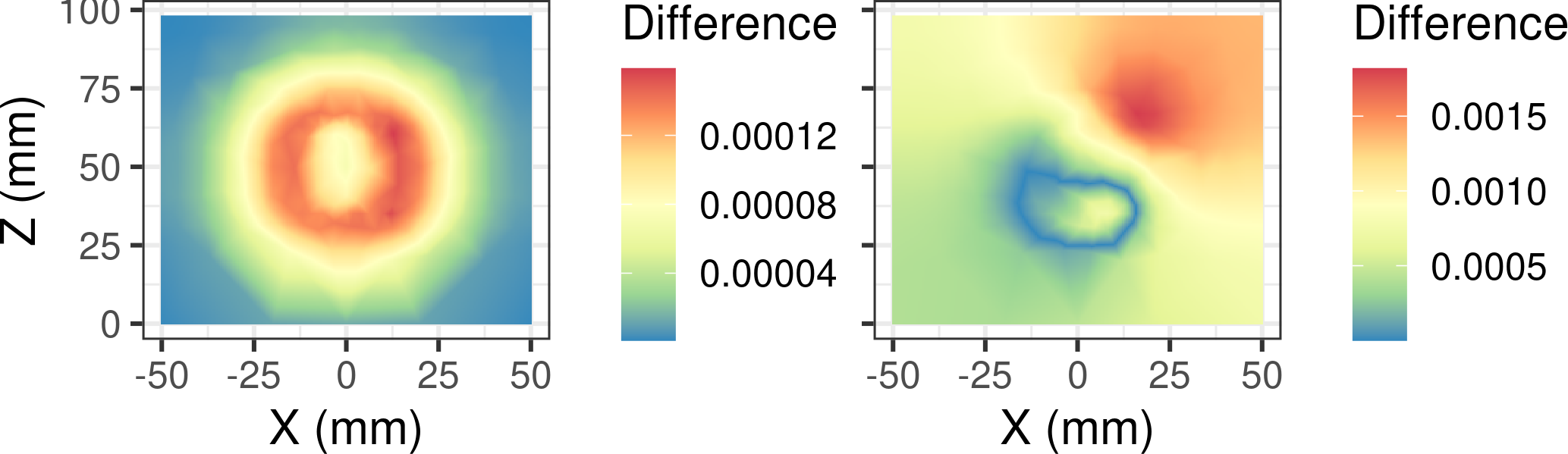}
  \caption{Numerical differences of MAGMA with lower tolerance for temperature (left) and for voltage (right).}  
  \label{fig:magma_low_tol_diffs}
\end{figure}

We note that, as the PSNR value indicated, the \texttt{MAGMA} solution is much
closer to the original than the others, for both temperature and
voltage. The differences increase from \texttt{cuSOLVER} to \texttt{QRMumps}, but all
solvers present the differences around the same area. The maximum
absolute difference values are still low, for example, 4e-6 in the
\texttt{QRMumps} voltage. However, for the \texttt{MAGMA\_low\_tol} case, in
\autoref{fig:magma_low_tol_diffs} we have a much bigger scale, and the
area where the numerical differences appear seems to be wider than the
others. For this configuration of the GMRES(\(m\)) solver, we could
perhaps start noticing visual differences in the final results
presentation, which may invalidate the simulation procedure. 

\section{Performance Evaluation}
\label{sec:org01eb23b}
\begin{figure}[!htbp]
  \centering
  \includegraphics[width=.48\textwidth]{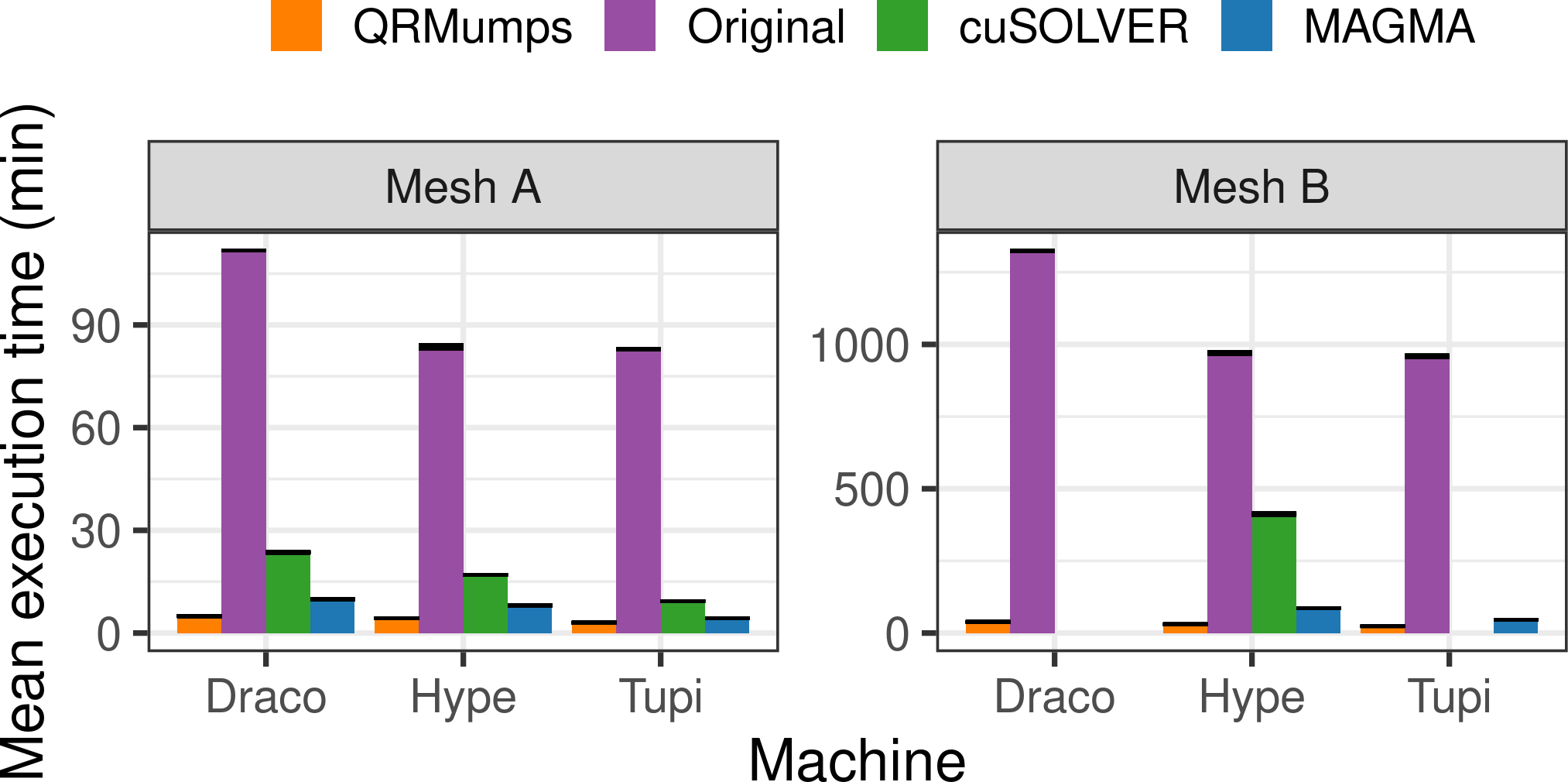}
  \caption{Makespan for the different machines and solvers.} 
  \label{fig:makespan}
\end{figure}

We conducted experiments with different machines and solvers,
collecting the total application execution
time. \autoref{fig:makespan} presents the overview of the execution
time by machine and solver, considering a confidence level of 99.7\%
after checking that the collected data is normality distributed. We
can observe that the \texttt{QRMumps} solver was faster than the others for all
machines and workloads, but the time difference between the GPU
solvers and \texttt{QRMumps} is reduced when we have a more powerful GPU in the
Tupi machine. We also notice that the sequential version is
significantly faster for the Hype and Tupi machines than in Draco, due
to the faster CPUs.

\begin{figure}[!htbp]
  \centering
  \includegraphics[width=.48\textwidth]{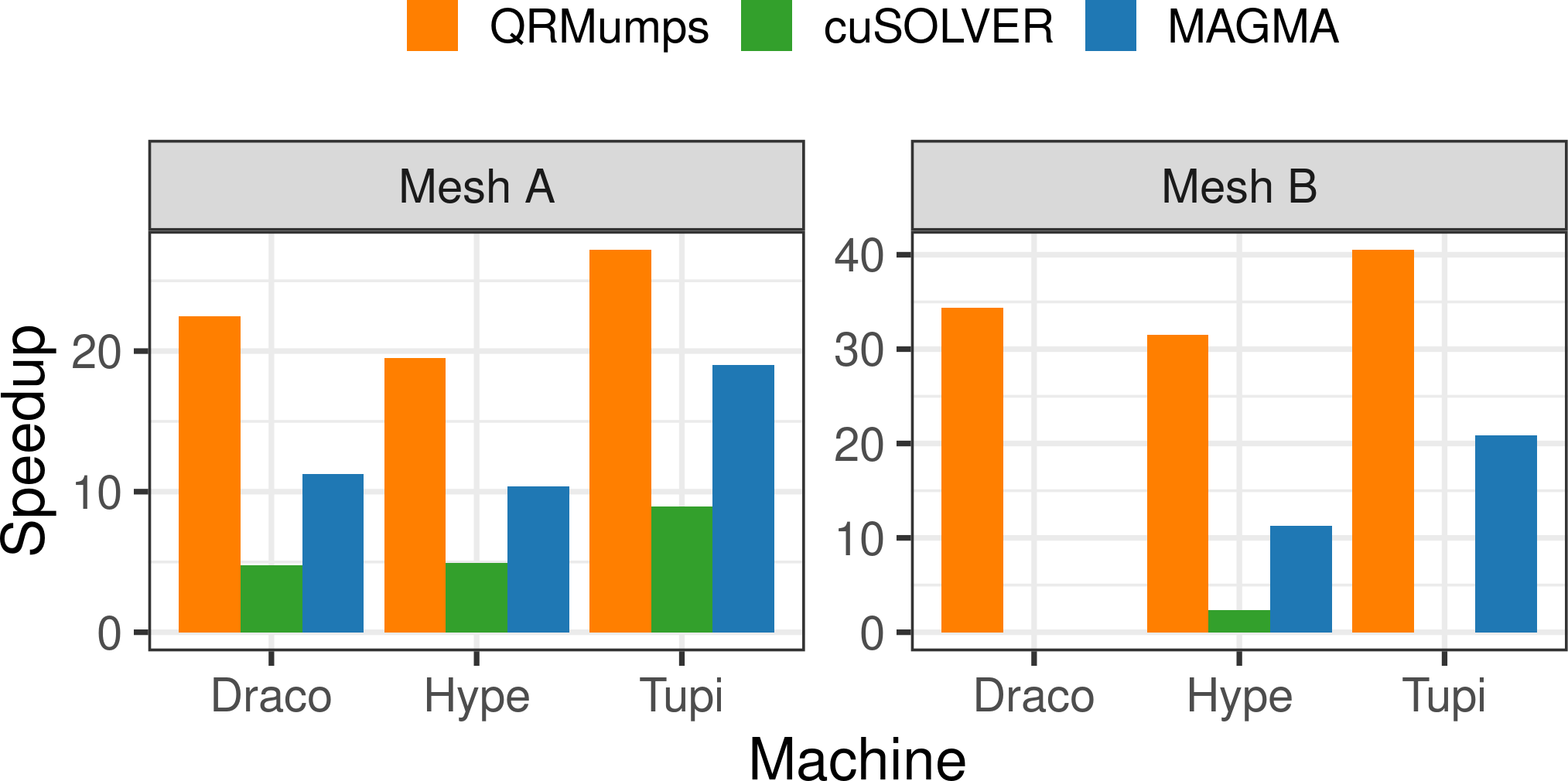}
  \caption{Speedup values for the different machines and solvers.} 
  \label{fig:speedup}
\end{figure}

We present the speedup metric in \autoref{fig:speedup}, where we can
better observe the effect of more powerful GPUs in the application
acceleration, principally among the GPU solvers. However, when we
focus on the different workloads, we observe a lower speedup for
\texttt{cuSOLVER} for the bigger input mesh B. This \texttt{cuSOLVER} results for the
mesh B suggest that its QR solver does scale well for bigger
problems. \texttt{MAGMA} and \texttt{QRMumps} presented a higher speedup for mesh B, but
the \texttt{MAGMA} increase was around one or two times faster.

In contrast, \texttt{QRMumps} presented increases around 13 times in the
speedup for mesh B, reaching up to 40 times of speedup compared to the
sequential version in the Tupi machine. However, for the \texttt{QRMumps}
solver, we noticed that although the original sequential version
makespan is very similar for Hype (966 min) and Tupi (956 min), the
speedup in the Tupi machine is higher than in the Hype machine, which
has more than twice the number of CPU cores than Tupi. This difference
may suggest that there is not enough parallelism in the matrix
factorization. Probable causes of this lack of parallelism might come
from the matrix structure itself and the configuration parameters used
in the \texttt{QRMumps} solvers. Thus, further optimizations can explore better
parameter values for these specific cases.

\begin{figure}[!htbp]
  \centering
  \includegraphics[width=.484\textwidth]{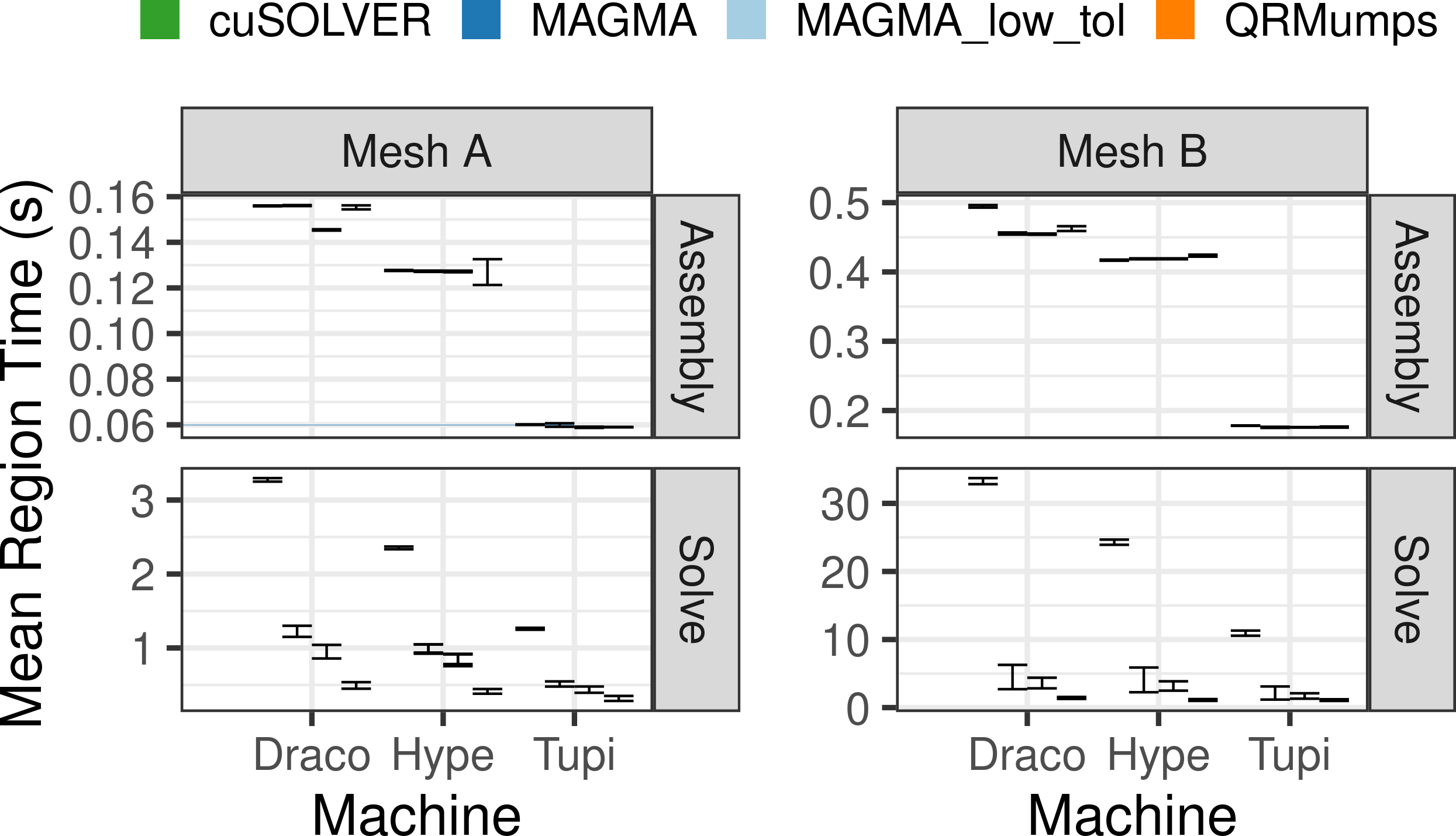}
  \caption{Trace iteration time for different assembly and solve regions.} 
  \label{fig:trace_results}
\end{figure}

A part of this speedup comes from the assembly step, where better GPUs
will run faster because it is a well scalable problem. We can observe
in \autoref{fig:trace_results} the differences in the mean duration of
the traced regions between all step iterations. This figure also
included the version \texttt{MAGMA\_low\_tol} to represent how lowering the
tolerance in the solve step for the GMRES(\(m\)) iterative solver
affects its duration. Nonetheless, we report that lowering the
tolerance to \(10^{-6}\) can reduce the total computation time by up
to 20\%. However, the reduction was less significant for some cases
because the application needed to execute more correction steps to
reach the desired convergence threshold value. Moreover, the tolerance
reduction may cause terrible effects in the numerical results, as
presented in \autoref{fig:magma_low_tol_diffs}. We also point out that
since we reduced the solve step time dramatically, memory copies take
a significant amount of time of the iteration, as much as the entire
assembly step. We can further optimize this step by using asynchronous
copies and overlapping them with computation. If we manage to do this
correctly, we can save up the memory copy time multiplied by the
number of iterations of our application execution. 

\section{Conclusion}
\label{sec:orgbadce78}
In this work, we presented the RAFEM application and the acceleration
of its most costly phase, using parallel sparse solvers, exploring
both iterative and direct solvers on multicore systems and GPU
accelerators. The solvers provided speedup values ranging from 2.12
times to 40.4 times (See \autoref{fig:speedup} in Section
\ref{sec:org01eb23b}), where we noticed a lack of scalability
in \texttt{cuSOLVER}, for which the speedup factor was lower for the bigger
workload case. This scalability issue also seems to be present for
this specific problem structure, where the \texttt{QRMumps} application
presented lower speedup value for the Hype machine, which has more CPU
cores than the Tupi machine, where the highest speedup value
occurred. This lower value may indicate a subtilization of the
computational resources, demanding an in-depth analysis to understand
better what happened. We also observed the beneficial effect of newer
and more powerful GPU hardware on the application
acceleration. Another discussion point involves iterative solvers
tolerance, which directly affects the number of iterations needed to
complete the solve step. We noticed in our experiments a reduction of
the execution time when we reduced the GMRES(\(m\)) tolerance from
\(10^{-10}\) to \(10^{-6}\). However, some cases presented a minimal reduction
because the RAFEM application corrector step executed more times to
reach the desired convergence threshold. Thus, there is a clear
trade-off between the solution numerical quality and the time needed
by the iterative solver, but this also affects the number of needed
corrector step iterations, where the exact relationship remains
unclear for the authors. However, early experiments point out that the
workload and application parameters play a role in this
matter. Moreover, when using lower tolerance, the simulated values
presented way more significant differences than the other solvers for
mesh A as presented in \ref{sec:orgc8cbffb}, which can be harmful
to the post-processing analysis of the results.

Lastly, future work can explore possible optimizations that can
accelerate the application as the memory transfers overlap with
computation and a more in-depth investigation in the \texttt{QRMumps} solver to
help in the fine-tuning of its parameters. Finally, undoubtedly more
performance can be achieved by improving the assembly stage,
considering one of the many newer proposed techniques discussed in
this work. The benefits of improving the assembly stage are
particularly crucial for more complex geometric shapes of elements and
a higher number of degrees of freedom. However, they could accelerate
the RAFEM application even more.   
\medskip\noindent
\textbf{Software and Data Availability}.  We endeavor to make our analysis
reproducible.  A public companion hosted at {\scriptsize\url{https://gitlab.com/mmiletto/hpcs2020}} contains the
source code, datasets, and instructions to reproduce our results.  A
perennial archive is also available in Zenodo at {\scriptsize\url{https://zenodo.org/record/4006573}}.

\section*{Acknowledgements}
We thank the following sponsors: FAPERGS MultiGPU (16/354-8) and
GreenCloud (16/488-9), the CAPES/Brafitec EcoSud 182/15, the Council
for Scientific and Technological Development (CNPq) under grant no
131347/2019-5 to the first author, and the CAPES/Cofecub 899/18.  This
study was financed in part by the Coordenação de Aperfeiçoamento de
Pessoal de Nível Superior - Brasil (CAPES) - Finance
Code 001. Experiments in this work used the PCAD infrastructure,
\url{http://gppd-hpc.inf.ufrgs.br}, at INF/UFRGS.

\bibliographystyle{IEEEtran}
\bibliography{refs}
\end{document}